\documentclass[preprint,12pt]{elsarticle}

\usepackage{graphicx}
\usepackage{amssymb}

\usepackage{lineno}
\usepackage{appendix}
\usepackage{amsmath}
\usepackage{amssymb}
\usepackage{multirow}
\usepackage{longtable}
\usepackage{algorithm}
\usepackage{algpseudocode}
\usepackage{booktabs}
\usepackage{geometry}
\usepackage{color}

\journal{Physica A: Statistical Mechanics and its Applications}

\makeatletter
\newcommand{\figcaption}{\def\@captype{figure}\caption}
\newcommand{\tabcaption}{\def\@captype{table}\caption}
\makeatother

\begin{document}

\begin{frontmatter}
\title{Will the Technological Singularity Come Soon? Modeling the Dynamics of Artificial Intelligence Development via Multi-Logistic Growth Process}
\author{Guangyin Jin\textsuperscript{1}\corref{cor1}}
\author{Xiaohan Ni\textsuperscript{2}\corref{cor2}}
\author{Kun Wei\textsuperscript{1}}
\author{Jie Zhao\textsuperscript{1}}
\author{Haoming Zhang\textsuperscript{3}}
\author{Leiming Jia\textsuperscript{1}}
\cortext[cor1]{Corresponding author: jinguangyin18@alumni.nudt.edu.cn}
\cortext[cor2]{The author contributed equally to this work}
\address{\textsuperscript{1} National Innovative Institude of Defense Technology, Beijing, China.}
\address{\textsuperscript{2} Capital Normal University, Beijing, China}
\address{\textsuperscript{3} Northwest Institute of Nuclear Technology, Xian, China}
\begin{abstract}
We are currently in an era of escalating technological complexity and profound societal transformations, where artificial intelligence (AI) technologies exemplified by large language models (LLMs) have reignited discussions on the ‘Technological Singularity’. ‘Technological Singularity’ is a philosophical concept referring to an irreversible and profound transformation that occurs when AI capabilities surpass those of humans comprehensively. However, quantitative modeling and analysis of the historical evolution and future trends of AI technologies remain scarce, failing to substantiate the singularity hypothesis adequately. This paper hypothesizes that the development of AI technologies could be characterized by the superposition of multiple logistic growth processes. To explore this hypothesis, we propose a multi-logistic growth process model and validate it using two real-world datasets: AI Historical Statistics and Arxiv AI Papers. Our analysis of the AI Historical Statistics dataset assesses the effectiveness of the multi-logistic model and evaluates the current and future trends in AI technology development. Additionally, cross-validation experiments on the Arxiv AI Paper, GPU Transistor and Internet User dataset enhance the robustness of our conclusions derived from the AI Historical Statistics dataset. The experimental results reveal that around 2024 marks the fastest point of the current AI wave, and the deep learning-based AI technologies are projected to decline around 2035-2040 if no fundamental technological innovation emerges. Consequently, the technological singularity appears unlikely to arrive in the foreseeable future. 
\end{abstract}
\end{frontmatter}

\section{Introduction}
\begin{figure}[htb]
\centering
\includegraphics[width=1 \textwidth]{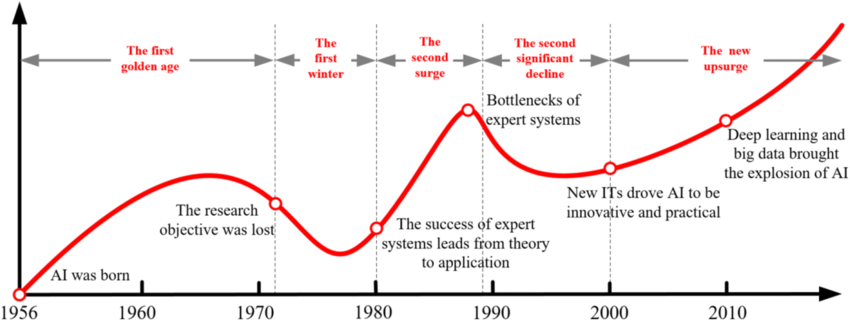}
\caption{AI technology has experienced three peaks and two troughs in history~\cite{wang2021artificial}.}
\label{fig:ai_his} 
\end{figure}

We are in an era of technological explosion, where emerging technologies are proliferating at an unprecedented pace, profoundly impacting the global socio-economic landscape, industries, and cognitive paradigms. Among these technologies, Artificial Intelligence (AI) stands out as particularly transformative, it has caused a stronger impact in society and its popularity has been increasing since 1986~\cite{fast2017long}.
AI has a history spanning nearly 70 years, with its conceptual foundations laid at the Dartmouth Conference in 1956~\cite{moor2006dartmouth}. 
Throughout this period, AI development has witnessed 'three peaks and two troughs', as shown in Fig~\ref{fig:ai_his}, and we are presently in the third wave, characterized by the 'Deep Learning' era. Deep learning, a method adept at uncovering hidden patterns in large datasets and solving practical problems, has significantly influenced the global industrial chain. However, it has also inevitably been over-hyped by some media and capital. Therefore, it is crucial to quantitatively model the historical development of AI technology and forecast its future trends. Such an approach allows us to comprehend the objective laws governing AI technology evolution and to evaluate its societal impact with greater rationality and composure.

\begin{figure}[htb]
\centering
\includegraphics[width=0.9 \textwidth]{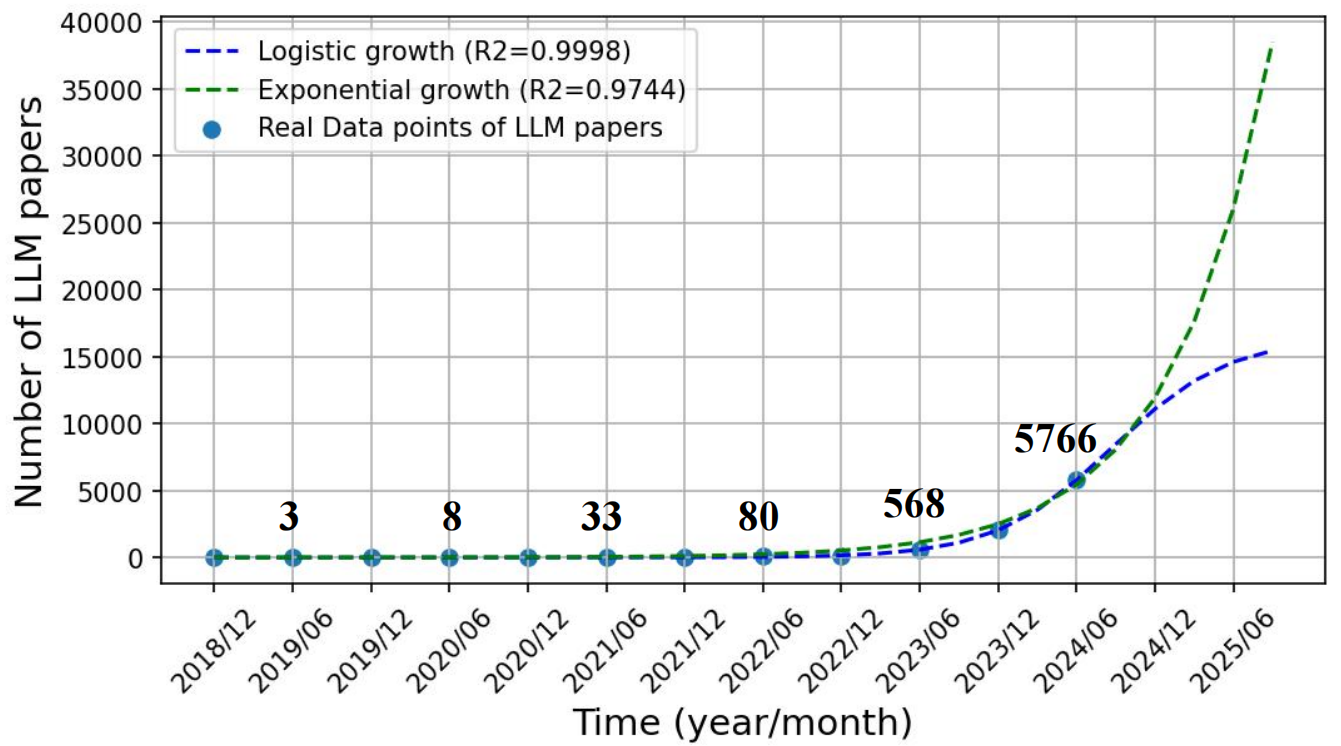}
\caption{The trends and fitting curves of the cumulative numbers of arXiv papers that contain the keyword 'large language model' in titles and abstracts.}
\label{fig:intro} 
\end{figure}

Since 2020, Large Language Models (LLMs) exemplified by the GPT series have emerged prominently~\cite{chang2024survey}, with the annual proliferation of notable LLM developments depicted explosively, as illustrated in Fig~\ref{fig:intro}. LLMs demonstrate remarkable capabilities in comprehending text, images, sounds, and even videos within the human domain, proficiently generating samples indistinguishable from ground truths.~\cite{liu2024sora,liu2023gpt}. 
Notably, GPT-4 recently passed the medical license examination~\cite{abd2023large}, prompting some researchers to speculate that it may have surpassed the 'Turing Test'~\cite{biever2023chatgpt}. These achievements underscore the growing belief among the public that the 'technological singularity' is drawing nearer. 
The technological singularity refers to the critical point at which the emergence of superintelligent AI drives an 'intelligence explosion', meaning that the development speed of artificial intelligence systems continues to grow at an infinite exponential rate~\cite{kurzweil2005singularity}.
Nevertheless, as researchers in the AI research community, it is imperative to recognize that we are still in the third wave of AI technology, nearing its zenith due to advancements such as LLMs. Despite these strides, LLMs remain extensions of classic deep learning architectures like Transformers~\cite{vaswani2017attention} and BERT~\cite{kenton2019bert}, lacking significant scientific theoretical breakthroughs.
Moreover, they exhibit several unresolved limitations such as hallucinations and high computational overhead~\cite{hadi2023survey,jin2023spatio,jin2023con}. They do not establish a complete understanding of the physical world but only mechanically summarize knowledge from massive data samples, rendering them less efficient in learning from sparse data.

Reflecting on the history of AI development, discussions about the technological singularity have been persistent, recurring with each wave of AI advancements. As early as 1965, Good~\cite{good1966speculations} posited that the AI singularity could likely arrive in the 20th century. Vinge~\cite{vinge1993coming} predicted that machines would surpass human intelligence between 2005 and 2030, while Yudkowsky~\cite{yudkowsky1996staring} forecasted the arrival of the AI singularity in 2021. Kurzweil~\cite{kurzweil2005singularity} anticipated that human-level AI would emerge around 2029, with the singularity occurring in 2045. Conversely, other scholars have expressed skepticism about the imminence of the technological singularity. In 2017, an email survey of authors who published papers at the NeurIPS and ICML conferences revealed that nearly half of the respondents doubted the likelihood of the AI singularity occurring in the foreseeable future~\cite{grace2018will}.
In summary, experts hold diverse opinions on the future trajectory of AI technology. However, there remains a notable absence of widely accepted and effective quantitative methods to predict the future of AI. Particularly lacking are methods to model the historical development of AI and make reliable extrapolations about its future trends.

To address the aforementioned problems, we need to propose an effective quantitative methods that can be expressed mathematically to characterize the dynamics of AI technology development.
As shown in Fig~\ref{fig:intro}, we fit the cumulative number of LLM-related papers on the Arxiv website using both logistic growth and exponential growth process. The logistic growth model aligns with past empirical predictions related to industrial or information technology development~\cite{harris2018logistic,burg2021moore}, while the exponential growth model corresponds to the viewpoint of 'technological singularity'~\cite{kurzweil2005singularity}.
From the experimental results, the logistic growth process exhibit a higher R-squared value compared to exponential growth, indicating a better fitting performance to the real data points. Considering also the historical pattern of AI technology development characterized by ‘three peaks and two troughs’, it is evident that AI development cannot follow the exponential growth. Therefore, we hypothesize that the development dynamics of AI technology may be modeled as the superposition of multiple logistic growth processes. Building on this hypothesis, we propose the multi-logistic model to fit the annual cumulative numbers of famous AI systems in the AI Historical Statistics dataset. Our model significantly outperformed other models, thus providing preliminary validation for our hypothesis.
We then conduct a comprehensive analysis of the parameters and derivative characteristics of our proposed model to forecast future trends in AI technology. Additionally, we focus on the current third wave of AI, conducting cross-validation experiments using the AI Arxiv Paper dataset.
The results from the cross-validation experiments align closely with our findings from the AI Historical Statistics dataset: the fast point of the current AI wave is anticipated around 2024, but without further theoretical breakthroughs, this third wave is likely to fade away around 2035-2040. 
Our main contributions in this paper are summarized as follows:
\begin{itemize}
\item We first propose the multi-logistic growth process to quantitatively model the dynamics of AI technology development and demonstrate its superior performance by comparative experiments on AI Historical Statistics dataset.
\item We conduct a comprehensive analysis of our proposed model and forecast the future development trends of AI technology.
\item We conduct cross-validation experiments on AI Arxiv Papers dataset to further verify the reliability of the prediction results.
\item Based on the experimental results of this paper, we reveal that the multi-logistic growth process could be an objective law of AI technology development, and provide some suggestions on how to treat the current development of AI technology more rationally and calmly.
\end{itemize}

The rest of this paper is organized as follows. We first sort out the related literature in Section 2. Then the hypothesis and methodology of modeling AI development dynamics is presented in Section 3. In Section 4, we conduct comparison experiment to evaluate our proposed model on real-world datasets  as well as the parameter and visualization analysis. In Section 5, we discuss and look into the future of AI development based on the experimental results. Finally, we conclude our contributions and future directions in Section 6. 

\section{Related work}

In this section, we review several related works on AI technology forecasting and logistic growth process respectively. 

\subsection{AI Technology Forecasting}
Although AI technology has developed so rapidly in the past decade, there is a relative lack of work on forecasting future AI technology trends. Currently, the relevant work on this topic can be roughly divided into three categories: literature mining-based, expert survey-based, and statistical modeling-based. 
The literature mining-based methods generally utilize various machine learning methods to capture key topics and keywords from a period of literature, thereby analyzing and forecasting the dynamics of different AI research fields. Based on the AMiner dataset, Shao et al~\cite{shao2022tracing} combined traditional literature review and bibliometric methods to further summarize the evolution of AI in the past decade from the development of connectionism and discuss their trends in the next decade. Dwivedi et al~\cite{dwivedi2023evolution} employed structural topic modeling (STM) to extract and visualize latent topics from AI research literature and analyzed their future trends.
The expert survey-based methods collect various opinions through questionnaires of community experts and conduct statistical analysis to forecast technology trends. Baum et al~\cite{baum2011long} presented an assessment of expert opinions regarding human-level AI research and analyzed when the Artificial General Intelligence (AGI) era will arrive. Gruetzemacher et al~\cite{gruetzemacher2021forecasting} used quantitative expert survey data to reveal that conference attendance has a statistically significant impact on all predictions. Halal et al~\cite{halal2013forecasting} proposed TechCast, an online system that pools background trends and the judgment of experts around the world, to forecast breakthroughs in frontier technology fields including AI technology.
The statistical modeling-based methods are to directly collect industry-related trend data and build models for fitting and prediction. Villalobos et al~\cite{villalobos2022will} established a growth model to analyze and  predict when the amount of data available for deep learning training will be exhausted. Modis~\cite{modis2022links} modeled 28 historical milestones by the logistic model and estimated the emergence time of future milestones. 
In general, the methods mentioned above have predicted and discussed the future of AI technology from a certain perspective, but there is still a lack of quantitative modeling and prediction work based on the global history of AI development.

\subsection{Logistic Growth Process}
The logistic growth process is a continuously differentiable function with an approximate 'S' shape, which was first proposed in population research filed~\cite{verhulst1845recherches}. 
Over the past half-century, the logistic function has been increasingly used in various academic fields except demography. In ecology, the growth of bacteria~\cite{mckendrick1912xlv} and crop yield~\cite{sepaskhah2011logistic} can be modeled by logistic function with corresponding growth factor. In the medical field, logistic functions have been widely used in tumor growth modeling and epidemic modeling. Laird~\cite{laird1964dynamics} proposed to employ Gompertz function~\cite{kirkwood2015deciphering}, a variant of logistic function, to fit the data of growth of tumors. Richards growth curve~\cite{richards1959flexible}, a flexible form of logistic function, has been successfully adopted to model the early phase of the COVID-19 outbreak worldwide~\cite{lee2020estimation}. Recently, Levene~\cite{levene2022skew} also proposed a skew logistic distribution for modeling COVID-19 waves.  
In sociology, logistic functions are often adopted to characterize socioeconomic dynamics and technological innovation. Carlota Perez~\cite{perez2003technological} employed logistic curves to model the Kondratiev cycles of economic dynamics. 
Arnulf Gruebler~\cite{grubler1990rise} has studied the diffusion of infrastructure such as canals, railways, motorways, and airlines in-depth and found evidence that their spread patterns follow the logistic growth process. Harris et al~\cite{harris2018logistic} used the logistic growth curve to model and forecast the production and consumption of US Energy.  Burg et al~\cite{burg2021moore} revisited Moore’s Law by establishing a logistic model of Intel chip density. 
In general, the logistic functions are applicable to model the groups with more complex components, which may have the capabilities to characterize the general evolution laws of many complex systems.

\section{Hypothesis and Methodology}
Some previous works proposed to use logistic growth process for modeling and simulating various technological developments such as mobile communication technology~\cite{kalem2021technology}, transportation technology~\cite{kucharavy2011logistic} and chip technology~\cite{burg2021moore}. The general form of the logistic function is formulated as follows:
\begin{equation}
L(t)=a_0+\frac{A}{1+\exp{\left(-\frac{t-M}{W}\right)}}
\end{equation}
where $L(t)$ denotes the value of the logistic function over time $t$, $a_0$ denotes the initial value, $A$ denotes the development ceiling. $M$ represents the midpoint of the logistic function, the point where the rate of development is fastest. $W$ is the parameter that controls the speed of logistic development. 

\begin{figure}[htb]
\centering
\includegraphics[width=1 \textwidth]{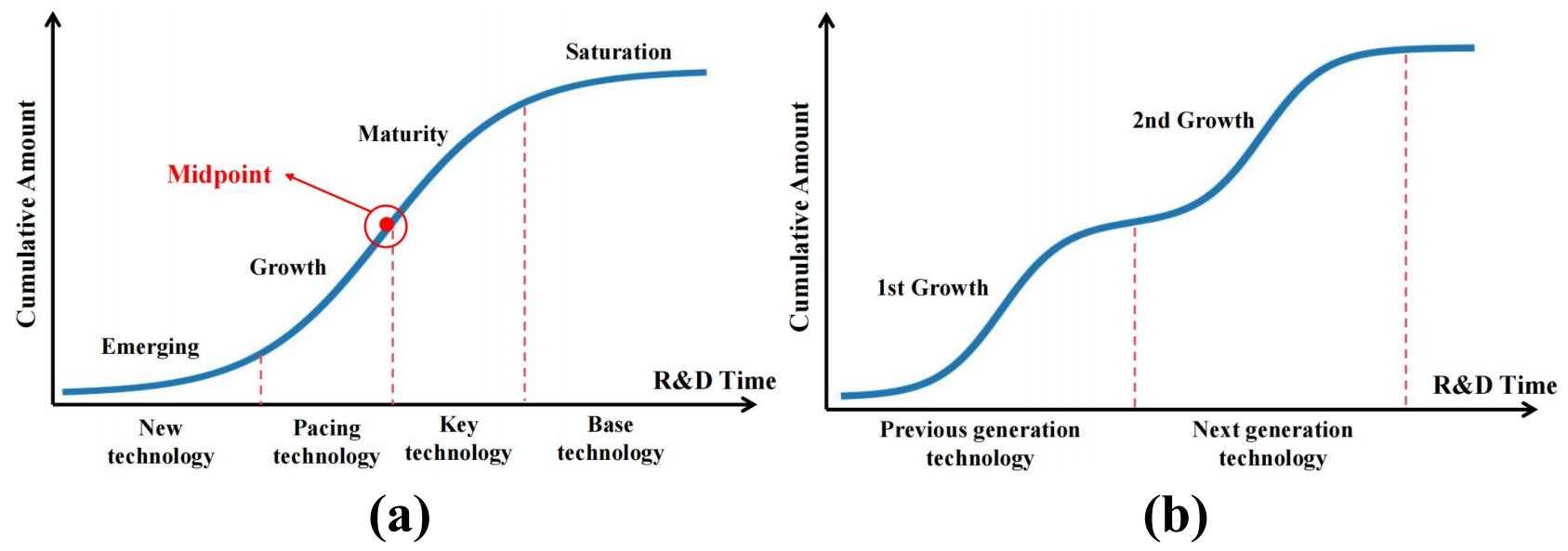}
\caption{The logistic growth and regrowth process of technological development.}
\label{fig:s_curve} 
\end{figure}

The growth of the logistic function can be divided into four stages, as shown in Figure~\ref{fig:s_curve}(a). The initial stage is similar to exponential growth, which represents the the emergence of new technologies. The growth rate in the second stage continues to increase but gradually transitions from exponential growth to approximately linear growth, which represents the technologies enter a period of pacing. The third stage is maturity, which which means that the technology has become a key technology but its development speed begins to decline. The final stage is saturation, which means that the technology has been transformed into a base technology and its development speed begins to decline exponentially. It should be noted that the midpoint is the critical point that distinguishes the growth stage and maturity stage of technological development. These four stages can be represented by variance intervals. The variance of the logistic function is calculated as follows:
\begin{equation}
\sigma = \frac{\pi\times W}{\sqrt{3}}
\label{eq:var}
\end{equation}
where emerging stage interval is $[M-2\sigma, M-\sigma]$, growth stage interval is $[M-\sigma, M]$, maturity stage interval is $[M, M+\sigma]$, and saturation stage interval is $[M+\sigma, M+2\sigma]$.


However, AI technology is a typical emerging technology that has experienced three ups and two downs in the past 70 years. The ordinary logistic growth process may not be able to accurately model its historical dynamics. 
\textbf{Therefore, we propose the hypothesis that the development of AI technologies may be the superposition of multiple logistic growth processes.} 
As shown in Figure~\ref{fig:s_curve}(b), the growth of the previous generation of technology has reached a bottleneck, and new scientific theoretical breakthroughs have led to the regrowth of the new generation of technology, and this cycle has constituted a multi-logistic function of technological development.


Based on the hypothesis that the development of AI technology could be a multi-logistic process, we propose a multi-logistic function to model the development history of AI. The formula is as follows:
\begin{equation}
L_n(t)=a_0+\sum_{i=1}^{N}\frac{a_i}{1+\exp{\left(-\frac{t-m_i}{w_i}\right)}}
\end{equation}
where $a_i$, $m_i$ and $w_i$ are respectively the development ceiling, time midpoint and speed control parameter of the $i_{th}$ logistic process.  

In this paper, we also pay attention to the first-order derivative of the multi-logistic process, where the formula for the first-order derivative is expressed as:
\begin{equation}
\frac{\mathrm{d}L_n(t)}{\mathrm{d}t}=\sum_{i=1}^{N}\frac{a_i\cdot\exp{\left(-\frac{t-m_i}{w_i}\right)}}{w_i\cdot(1+\exp{\left(-\frac{t-m_i}{w_i}\right)})^2}
\end{equation}
The first-order derivative represent the speed of AI technology development, which can also represent the heat of AI development. 


In order to solve the unknown parameters in the multi-logistic equation, we employ Levenberg-Marquardt algorithm~\cite{fletcher1971modified}, also known as the damped least-squares (DLS) method, which has been widely used for solving non-linear least squares problems. Given a set of data pairs $(t_i, y_i)$, our goal is to find the optimal parameters $\theta$ and minimize squares of the deviations $S(\boldsymbol{\theta})$: 
\begin{equation}
\hat{\boldsymbol{\theta}} \in \operatorname{argmin}_{\boldsymbol{\theta}} S(\boldsymbol{\theta}) = \operatorname{argmin}_{\boldsymbol{\theta}} \sum_{i=1}^m\left[y_i-L_n\left(t_i, \boldsymbol{\theta}\right)\right]^2
\end{equation}
Specifically, the parameters learning in DLS is an iterative process as follows:
\begin{align}
&\mathbf{J}_i=\frac{\partial L_{n}\left(t_i, \boldsymbol{\theta}\right)}{\partial \boldsymbol{\theta}},\\
&\mathbf{\delta}=\frac{\mathbf{J}^{\mathrm{T}}[\mathbf{Y}-\mathbf{L_n}(\boldsymbol{\theta})]}{\mathbf{J}^{\mathrm{T}} \mathbf{J}+\lambda \operatorname{diag}\left(\mathbf{J}^{\mathrm{T}} \mathbf{J}\right)},\\
&\boldsymbol{\theta}_{new}\rightarrow\boldsymbol{\theta}_{old}+\mathbf{\delta}
\end{align}
where $\mathbf{J}_i$ is the gradient of $L_{n}$ with respect to $\boldsymbol{\theta}$. $\mathbf{J}$ is the Jacobian matrix, whose $i_{th}$ row equals $\mathbf{J}_i$. $\mathbf{\delta}$ is the increment of the parameters $\boldsymbol{\theta}$ to be learned compared to the previous iteration, and $\lambda$ is the damping coefficient that can be adaptively adjusted during iteration. 

After obtaining the learned parameters of the multi-logistic growth model, it is necessary to calculate the confidence intervals to observe the uncertainty of the model predictions. Assuming the predicted mean of the model follows a normal distribution, the standard deviation of the prediction $\hat{Y}_t$ is calculated as follows:
\begin{align}
S_{\hat{Y}_t} &= \sqrt{\hat{\sigma}^2\left[\frac{1}{n}+\frac{\left(x_t-\bar{x}\right)^2}{\sum x_i^2}\right]}
\label{eq:confident}
\end{align}
where $\hat{\sigma}^2$ denotes the regression error variance, $n$ denotes the number of samples, $x_t$ denotes the independent variable at the current time step, $\bar{x}$ is the mean of all independent variables. The confidence interval can be obtained by $[\hat{Y}_t-t_{\alpha/2}(n-1)\cdot S_{\hat{Y}_t}, \hat{Y}_t+t_{\alpha/2}(n-1)\cdot S_{\hat{Y}_t}]$, where $t_{\alpha/2}(n-1)$ can be obtained by querying the t-distribution table.

\section{Experiment}
In this section, we conduct extensive experiments to demonstrate our proposed model on real-world datasets to answer the following research questions:
\begin{itemize}
\item \textbf{RQ1:} Is the model based on multi-logistic growth reasonable for modeling the development history of AI technology?
\item \textbf{RQ1:} How does our proposed multi-logistic growth model perform compared with other growth models in fitting the development trends of AI technology?
\item \textbf{RQ3:} Do the parameters learned from the proposed multi-logistic growth model have practical significance? What is the prediction of the learned model for future AI technology trends?
\item \textbf{RQ4:}  Could we get similar predictions or conclusions about the future trends of AI technology based on data from different sources?   
\end{itemize}

\subsection{Dataset and Metrics}
As shown in Table~\ref{tab:data}, we use two different datasets in this paper to verity our hypothesis that the development of AI technology follows a multi-logistic growth process. 
\begin{table}[h]
\caption{Dataset description and statistics.}
\centering
	\scalebox{1.0}{
	\begin{tabular}{cccc}
		\hline
		Datasets  & \#Subjects & \#TimeSteps &\#TimeRange \\ \hline
		AI Historical Statistics   & 4     & 60    & 1950 - 2023    \\
		AI Arxiv Paper    & 9     & 16  & 2008 - 2023  \\ 
            GPU Transistor    & 1     & 34  & 1982 - 2022  \\ 
            Internet User     & 1     & 28  & 1992 - 2020  \\ \hline
	\end{tabular}}
	\label{tab:data}
\end{table}

\begin{itemize}
    \item \textbf{AI Historical Statistics\footnote{https://ourworldindata.org/artificial-intelligence}:} This dataset records the cumulative number of famous systems developed in the field of AI each year since the 1950s. The dataset contains four major categories, namely total, academia, industry, and industry-academia collaboration. Our purpose in using this dataset is to explore the dynamics of the history of historical AI technology development by multi-logistic process.
    \item \textbf{AI Arxiv Paper:} This dataset records the cumulative number of papers related to the AI field published on Arxiv website\footnote{https://arxiv.org/} each year from 2008 to 2023. There are nine subjects in this dataset: Total, Computer Vision and Pattern Recognition (CV), Computation and Language (CL), Data Structures and Algorithms (DA), Human-Computer Interaction (HCI),  Software Engineering (SE), Robotics (RO), Cryptography and Security (CS),  and Machine Learning (ML). This dataset aims to cross-validate some of the conclusions obtained from AI Historical Statistics.
    \item \textbf{GPU Transistor\footnote{https://www.kaggle.com/datasets/maryanalyze/tables-on-transistor-count-wikipedia-page}:} This dataset records the annual changes in the number of transistors in newly developed GPUs from 1982 to 2022. We use the average number of transistors to measure the annual transistor count of newly developed GPUs, where a higher transistor count indicates stronger computational capability of the hardware.
    \item \textbf{Internet User\footnote{https://data.worldbank.org/indicator/IT.NET.USER.ZS}:} This dataset records the annual changes in the number of internet users from 1992 to 2020. An increasing number of internet users generates more data, including text, images, and audio, thereby providing richer training data for deep learning.
\end{itemize}

We read data and fit the model with LMFIT tool in Python 3.8 environment. 
In order to evaluate the fitting performance of different functions, we adopt four metrics in this paper: Reduced Chi-Square ($\chi^2_\nu$), Akaike Information Criterion (AIC), Mean Absolute Percentage Error (MAPE), and R-Square ($R^2$), which are respectively expressed as:
\begin{equation}
{\rm \chi^2_\nu}= \frac{1}{N-N_{\rm v}}{\sum_i^N (y_i-\hat{y}_i)^2} 
\end{equation}
\begin{equation}
{\rm AIC} =  N \ln(\frac{1}{N}\sum_i^N (y_i-\hat{y}_i)^2) + 2 N_{\rm v}
\end{equation}
\begin{equation}
{\rm MAPE} = \frac{1}{N}\sum_i^N \frac{|y-\hat{y}_i|}{y}
\end{equation}
\begin{equation}
{\rm R^2} =  1-\sum_i^N (y_i-\hat{y}_i)^2/\sum_i^N (y-\bar{Y})^2
\end{equation}
where $y_i$ denotes the ground truths, $\hat{y}_i$ denotes the estimated value, $\bar{Y}$ represents the mean of all dependent variables. $N$ and $N_{\rm v}$ are are the number of samples and the number of parameters respectively. Except for $R^2$, the lower the other three metrics, the better the fitting performance. 

\subsection{Segment Fitting Experiments (RQ1)}
\begin{figure}[htb]
\centering
\includegraphics[width=1 \textwidth]{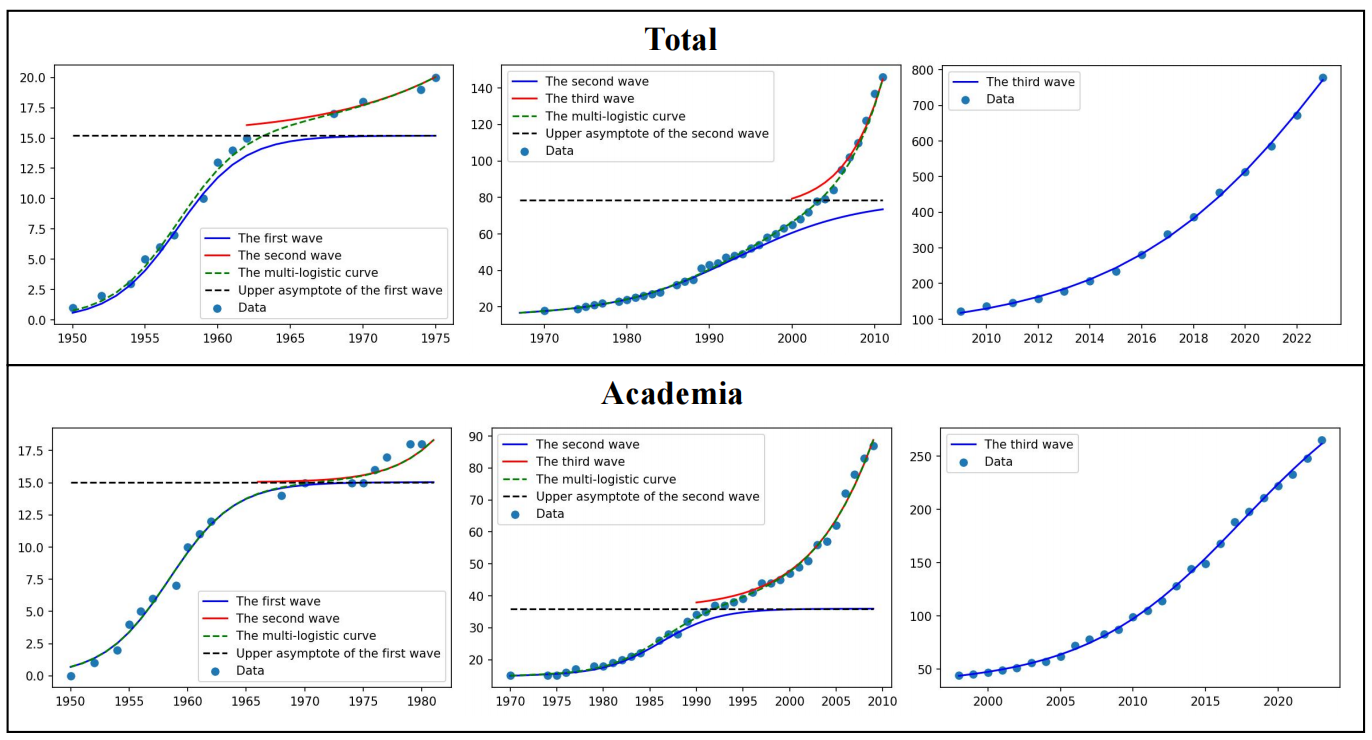}
\caption{The visualization of segmented fitting for different AI waves of overall and academia.}
\label{fig:fit_1} 
\end{figure}

\begin{figure}[htb]
\centering
\includegraphics[width=0.9 \textwidth]{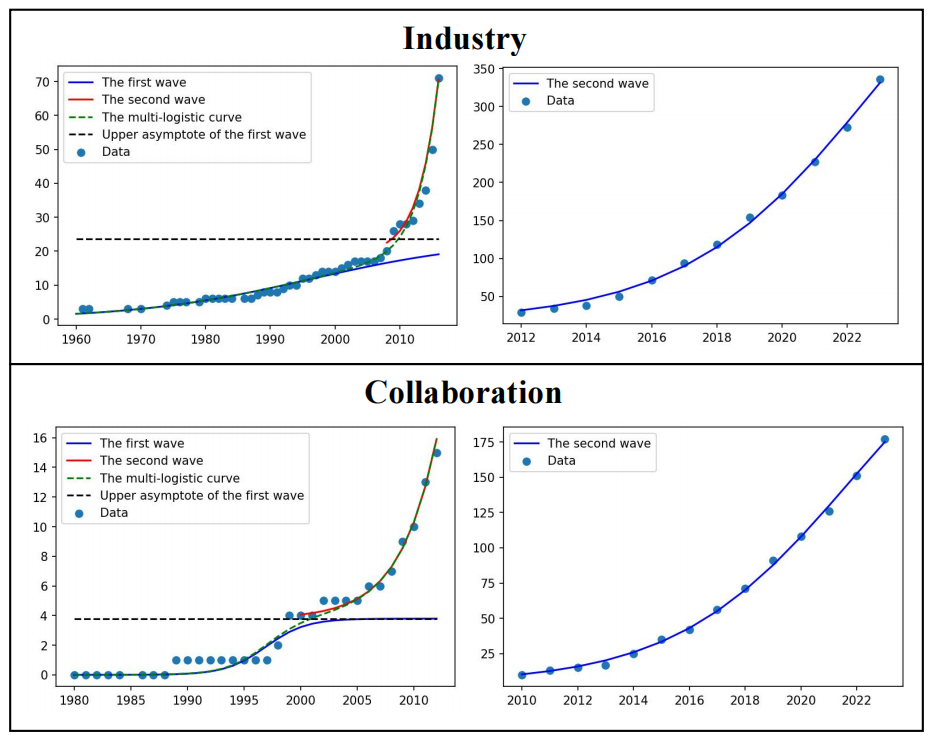}
\caption{The visualization of segmented fitting for different AI waves of industry and industry-academia collaboration.}
\label{fig:fit_2} 
\end{figure}

In order to illustrate the rationality of the multi-logistic growth model, we visualize its segmented fitting performance on four subjects of the AI Historical Statistics dataset in Fig~\ref{fig:fit_1} and Fig~\ref{fig:fit_2}.
Fig~\ref{fig:fit_1} shows the segmented fitting performance of overall and academia, while Fig~\ref{fig:fit_2} shows the segmented fitting performance of industry and industry-academia collaboration. Since industry only began to emerge in the second wave of AI, there are only two waves of industry and industry-academia collaboration.
No matter from Fig~\ref{fig:fit_1} or Fig~\ref{fig:fit_2}, it can be observed that for each AI wave, the corresponding segmented logistic growth curve fits the existing data points well. We also focus on showing the critical transitions of different segmented logistic growth curves, where the blue solid line represents the previous segment logistic growth curve, and the red solid line represents the next segment logistic growth curve. The black dashed line represents the upper asymptotic line on the previous segment logistic growth curve, and green dashed line represents multi-logistic growth curve. Some data points have poor accuracy in segmented fitting, and they happen to fall within in the critical transition interval where the next segment logistic growth curve begins to overlap. However, the multi-logistic growth curve precisely fits these data points more accurately.
From these visualizations, the rationality of the logistic growth model is verified, and it is proved that the multi-logistic growth model can indeed better fit the historical development trends of AI technology. 

\subsection{Comparative Experiments (RQ2)}
To verify that the proposed multi-logistic process model can better characterize the historical dynamics of AI technology development, we compare its fitting performance with three other commonly used mathematical models. The other four baselines are the Log Periodic Power Law (LPPL)~\cite{drozdz2003log}, the ordinary logistic model, the exponential growth model, and the polynomial model. As shown in Table~\ref{tab:comparative}, the experimental results on AI Historical Statistics dataset are displayed. Obviously, no matter which subject, the multi-logistic process model performs better in terms of $\chi^2_\nu$, AIC, MAPE and $R^2$. To be specific, the overall fitting performance of the exponential growth model is not only weaker than the proposed multi-logistic process model, but also weaker than the ordinary logistic model, which strongly refutes some previous views~\cite{kurzweil2001law,kurzweil2005singularity} that AI technology will continue to grow exponentially. 
Meanwhile, multi-logistic process model performs significantly better than logistic model, which means that the ordinary logistic process cannot accurately model the generational re-growth phenomenon in the development history of AI technology. In addition, judging from the absolute performance of the four indicators, the multi-logistic process model almost perfectly characterize the development history of AI technology.
Based on these evidences, we have preliminarily verified the correctness of the hypothesis that the development of AI technology follows a multi-logistic growth process.

\begin{table}[!htb]
\caption{Performance comparison of baseline models and multi-logistic process on AI Historical Statistics dataset. The best results are bolded while sub-optimal results are marked by the asterisk.}
\centering
\scalebox{1.0}{
	\begin{tabular}{clcccc}
		\hline
		Subject                 & Models    & $\chi^2_\nu$            & AIC         & MAPE   & $R^2$           \\ \hline
		\multirow{4}{*}{Total} & Multi-Logistic         & \textbf{14.46}          & \textbf{168.55}          & \textbf{0.0343}    & \textbf{0.9996}    \\
	& LPPL   & 15.45* & 168.88* & 0.1540* & 0.9995*\\	
        & Logistic   & 462.21 & 371.08 & 0.4975 & 0.9847*\\
		& Exponential    & 1600.71 & 443.68 & 0.3469 & 0.9452 \\
		& Polynomial  & 2322.37 & 467.94 & 1.2194  & 0.9231 \\ \hline
		\multirow{4}{*}{Academia}  & Multi-Logistic        & \textbf{3.70}          & \textbf{86.80}          & \textbf{0.0423}    & \textbf{0.9993}  \\
        & LPPL  & 7.22* & 156.55* & 0.0835* & 0.9979*\\
		& Logistic   & 47.97 & 235.16 & 0.2389 & 0.9905\\
		& Exponential    & 55.54 & 242.02 & 0.2676 & 0.9886 \\
		& Polynomial  & 60.62 & 249.20 & 0.3318  & 0.9881 \\ \hline
		\multirow{4}{*}{Industry}  & Multi-Logistic        & \textbf{6.25}          & \textbf{115.63}          & \textbf{0.0924}    & \textbf{0.9987}  \\
        & LPPL   & 6.41* & 120.41* & 0.1205* & 0.9987*\\
		& Logistic   & 54.18 & 242.46 & 0.5688 & 0.9883\\
		& Exponential    & 1206.56 & 426.72 & 0.9124 & 0.7325 \\
		& Polynomial  & 824.19 & 405.78 & 2.1946  & 0.8234 \\ \hline
		\multirow{4}{*}{Collaboration}  & Multi-Logistic        & \textbf{1.05}          & \textbf{8.72}    & \textbf{0.0762}    & \textbf{0.9993}  \\
	& LPPL  & 9.77* & 14.90* & 0.3305 & 0.9993*\\
        & Logistic   & 2.68* & 62.01 & 0.1694* & 0.9982\\
		& Exponential    & 490.13 & 372.67 & 4.0780 & 0.6635 \\
		& Polynomial  & 211.59 & 324.20 & 4.4485  & 0.8596 \\ \hline   
	\end{tabular}}
	\label{tab:comparative}
\end{table}


\begin{figure}[ht]
\centering
\includegraphics[width=1 \textwidth]{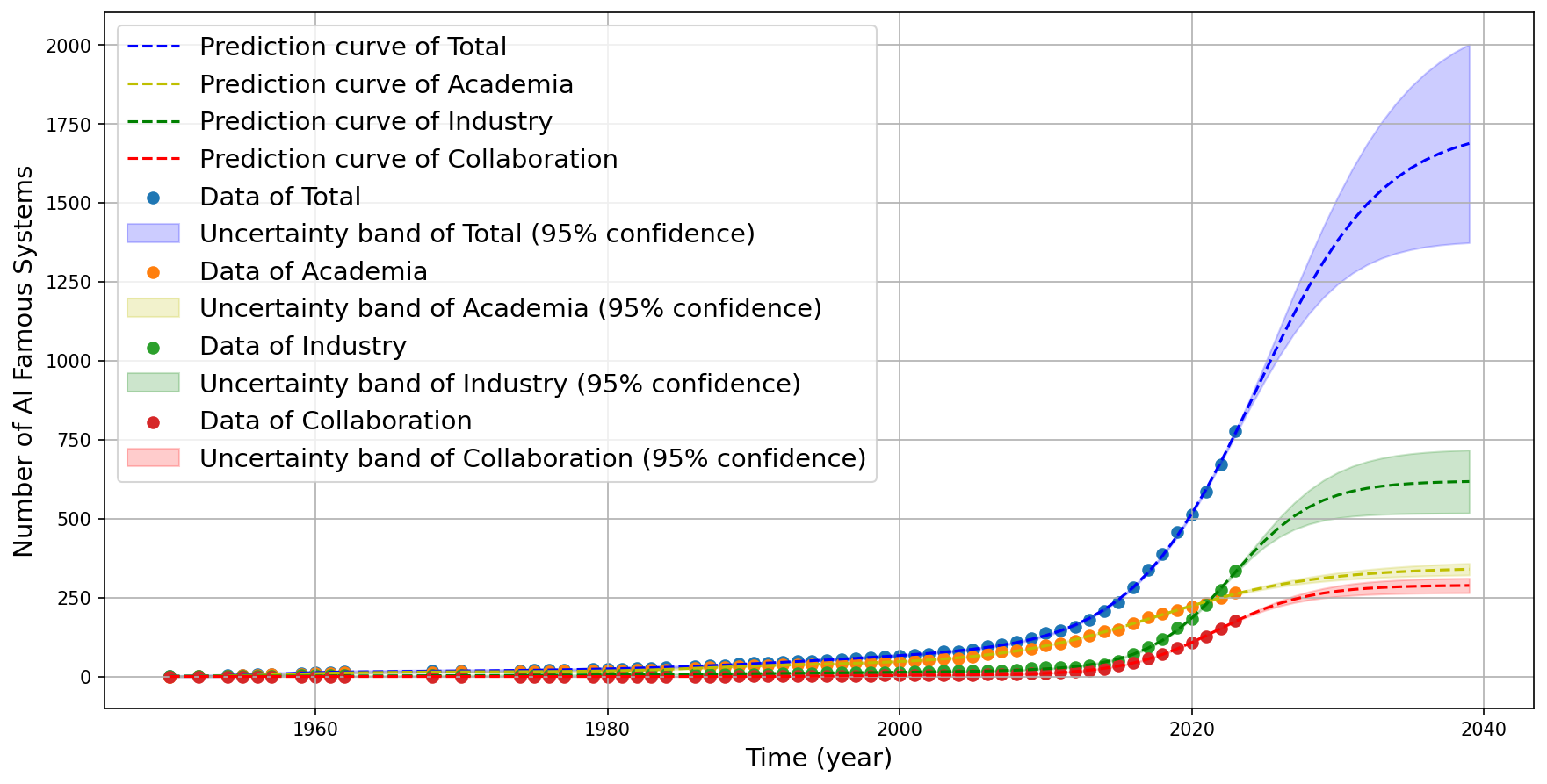}
\caption{The prediction of the number of famous AI systems with $95\%$ confidence intervals.}
\label{fig:prediction} 
\end{figure}

\begin{figure}[htb!]
\centering
\includegraphics[width=0.85 \textwidth]{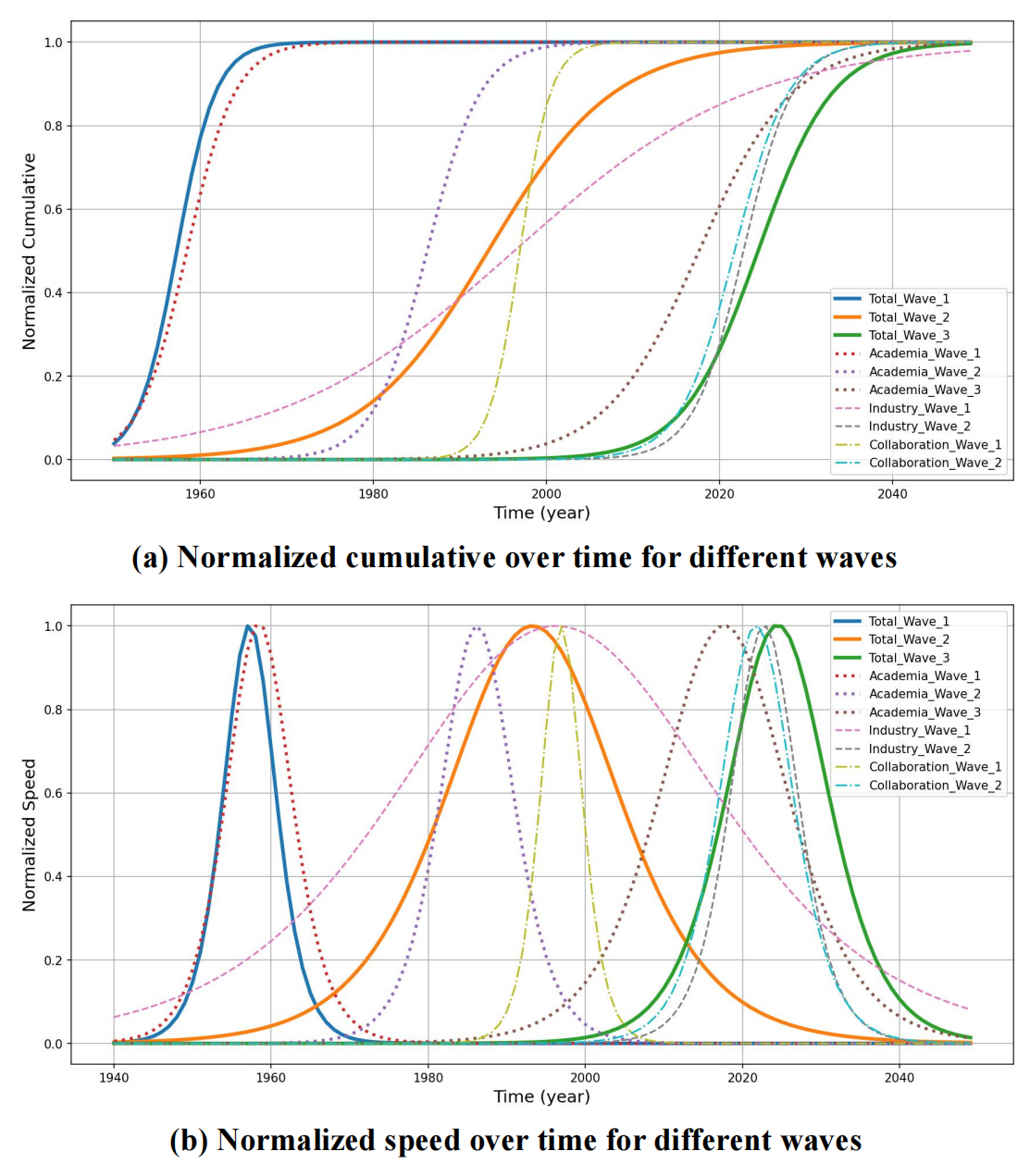}
\caption{The visualization of the attributes for different AI waves.}
\label{fig:three_wave} 
\end{figure}

\begin{table}[h]
\caption{The optimal parameters of multi-logistic function.}
\centering
	\scalebox{1.0}{
	\begin{tabular}{ccccc}
		\hline
		Parameters  & Total & Academia &Industry &Collaboration \\ \hline
		$a_1$   & 15.20    & 15.04    & $\times$   & $\times$\\
		$w_1$   & 2.25     & 2.78     & $\times$   & $\times$\\ 
            $m_1$   & 1957.25  & 1958.43  & $\times$   & $\times$\\ \hline  
		$a_2$   & 63.44    & 20.91    & 23.59   & 3.79\\
		$w_2$   & 7.32     & 3.09     & 13.66   & 1.77\\ 
            $m_2$   & 1993.25  & 1986.21  & 1996.28 & 1996.92\\ \hline       
		$a_3$   & 1664.85  & 309.70   & 596.68   & 285.43\\
		$w_3$   & 4.33     & 5.45     & 2.87   & 3.12\\ 
            $m_3$   & 2024.46  & 2017.63  & 2022.76 & 2021.73\\ \hline    
	\end{tabular}}
	\label{tab:multi-logistic parameters}
\end{table}

\begin{table}[h]
\caption{The calculation of variance ($\sigma$) and time range ($T$).}
\centering
	\scalebox{1.0}{
	\begin{tabular}{ccccc}
		\hline
		Variables  & Total & Academia &Industry &Collaboration \\ \hline
		$\sigma_1$   & 4.07    & 5.04    & $\times$   & $\times$\\
		$T_1(1\sigma_1)$   & 1953-1961     & 1953-1963     & $\times$   & $\times$\\ 
            $T_1(2\sigma_1)$   & 1950-1965  & 1950-1969  & $\times$   & $\times$\\ \hline  
		$\sigma_2$   & 13.28    & 5.61    & 24.78   & 3.22\\
		$T_2(1\sigma_2)$   & 1980-2007     & 1981-1992     & 1972-2021   & 1994-2000\\ 
            $T_2(2\sigma_2)$   & 1967-2020  & 1975-1997  & 1950-2046 & 1990-2003\\ \hline       
		$\sigma_3$   & 7.86  & 9.89   & 5.20   & 5.67\\
		$T_3(1\sigma_3)$   & 2017-2032     & 2008-2028     & 2018-2028   & 2016-2027\\ 
            $T_3(2\sigma_3)$   & 2009-2040  & 1998-2037  & 2012-2033 & 2010-2033\\ \hline    
	\end{tabular}}
	\label{tab:variance}
\end{table}

\subsection{Parameter and Visualization Analysis (RQ3)}
Next, we analyze the parameters of the multi-logistic process, further discuss their practical physical meanings, and visualize the prediction results and related curves of the model. As shown in Table~\ref{tab:multi-logistic parameters}, we display the optimal values of $a_i$, $w_i$ and $m_i$ in the multi-logistic growth process. It should be noted that the overall development of AI technology and the development of academia have experienced three waves, thus we employ three-order multi-logistic model, while the development of industry and academia-industry collaboration has only two waves (the first wave of AI in history was almost entirely confined to academia), so we adopt two-order multi-logistic process models for them. 
Based on the obtained parameters $a_i$ and $w_i$, we can compute the corresponding variances and stage intervals for different waves and subjects according to Eq.~\ref{eq:var}, which are shown in Table~\ref{tab:variance}.

To more intuitively present the modeling and forecasting of AI technology development based on multi-logistic growth process, we visualize the prediction curves with $95\%$ confidence interval uncertainty bands in Figure~\ref{fig:prediction}. 
The confidence intervals for the prediction curves are calculated using Eq.~\ref{eq:confident}.
From the prediction curves in Figure~\ref{fig:prediction}, we can intuitively find that in 2040, the total number of well-known AI systems may reach about twice that of 2023. In other words, the incremental AI development in the next less than 20 years will reach the cumulative level of the past 70 years.
Based on the given prediction curves, we also visualize the normalized decomposition cumulative curves and normalized first-order derivative curves, as shown in Figure~\ref{fig:three_wave}. 
Based on the comprehensive analysis of these charts, we summarized the following important findings:

\textbf{Industry is leading the current wave of AI technology:} 
The data presented in Table~\ref{tab:multi-logistic parameters} and Figure~\ref{fig:prediction} delineate a notable shift in the dynamics of AI technology development across different waves. Initially, academia held a dominant position during the first two waves of AI technology. However, in the ongoing third wave, there has been a perceptible transition with industry taking the forefront in driving overall technological progress.
This observation underscores the complementary roles played by academia and industry in the evolution of AI technology. Academic research serves as the bedrock during the nascent stages, laying down theoretical frameworks and fundamental innovations. Subsequently, industrialization plays a pivotal role in translating these theoretical advancements into tangible applications, thereby driving the continuous advancement of technology.

\textbf{The intergenerational development of AI technology is overlapping:} 
The evolution of AI technology spans across generations, characterized by overlapping phases as depicted in Figures~\ref{fig:three_wave}. Each wave begins accumulating momentum even before the preceding wave concludes, highlighting the seamless transition and advancement within the field. While the timing of the emergence of the fourth wave of AI remains uncertain, historical patterns suggest it will likely emerge quietly as the third wave is close to exhaustion. This observation underscores the persistent and evolving nature of technological advancement, where innovations build upon prior achievements to propel the field forward.

\textbf{The second wave of AI technology plays a connecting role:} 
The calculation results presented in Table~\ref{tab:variance}, along with the curves shown in Figure~\ref{fig:three_wave}, indicate that the second AI wave had the longest duration from its inception to its conclusion. This underscores its crucial role in bridging the past and future of AI development. Historically, this finding aligns with the trajectory of AI advancements. During the second AI wave, numerous theoretical models and intelligent systems emerged, which have proven foundational for the current deep learning era. For instance, back-propagation theory~\cite{rumelhart1986learning}, long short-term memory networks~\cite{hochreiter1997long}, and convolutional neural networks~\cite{lecun1995convolutional} were all developed during this period, laying the groundwork for modern AI. Additionally, IBM's Deep Blue~\cite{hsu1999ibm} became the first intelligent system to master complex human intellectual games, marking a significant milestone. Therefore, the second wave of AI continues to exert a profound influence on contemporary AI technologies, demonstrating its enduring impact and significance in the evolution of the field.

\textbf{AI technology is currently at its fastest development point:}
From the results in Table~\ref{tab:multi-logistic parameters} and Figure~\ref{fig:three_wave}(b), the projected midpoint of the ongoing third wave of AI is approximately 2024, suggesting we are currently experiencing the peak velocity of this technological surge. This observation helps elucidate the rapid expansion observed in Large Language Models (LLMs) presently. Nevertheless, as indicated by the predictive trend illustrated in Figure~\ref{fig:three_wave} and the temporal analysis detailed in Table~\ref{tab:variance}, absent any substantial new technological advancements, the current AI wave is anticipated to plateau between 2035 and 2040. 

\begin{figure}[htb]
\centering
\includegraphics[width=1 \textwidth]{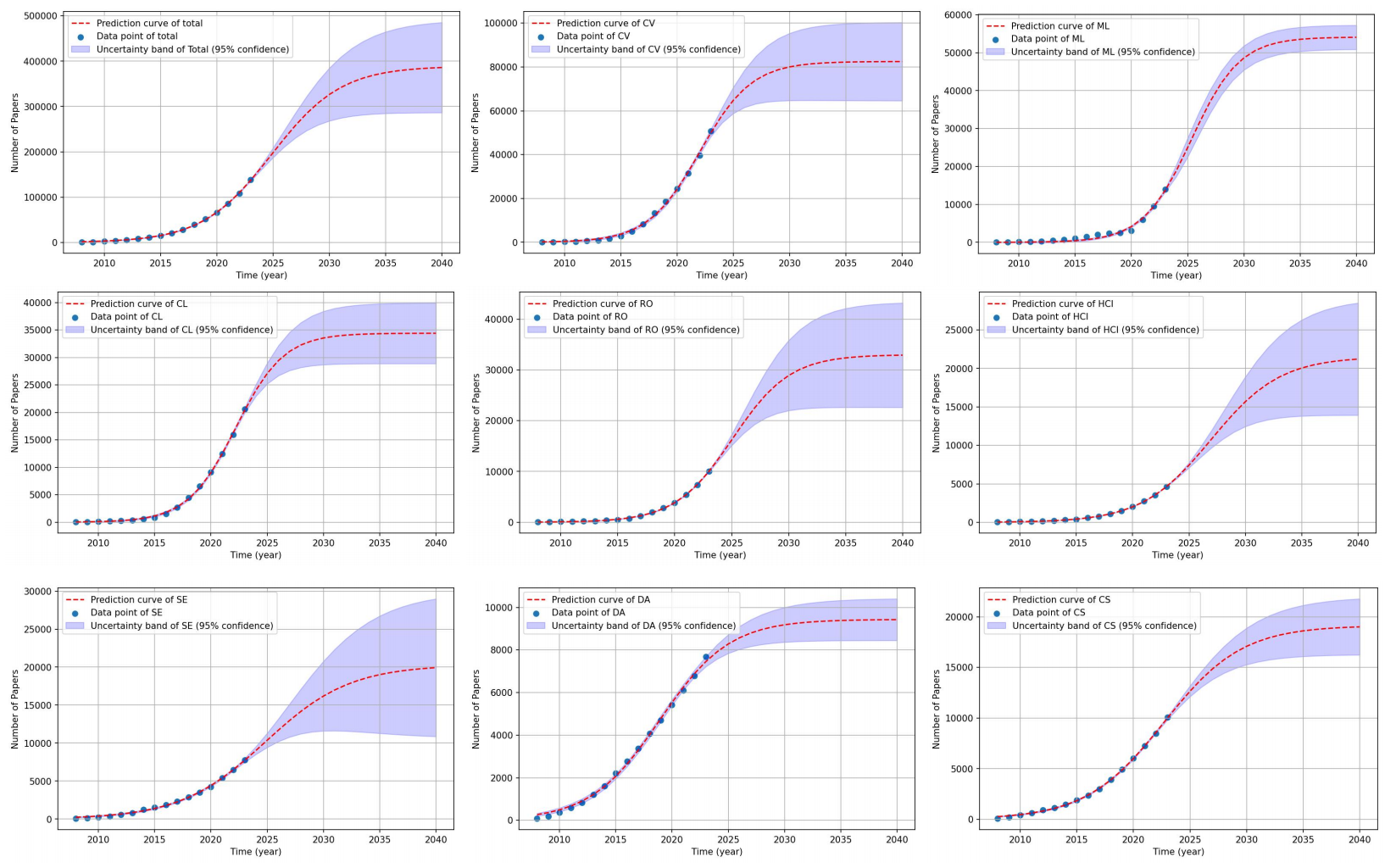}
\caption{The data and prediction results on the annual cumulative number of AI-related papers.}
\label{fig:arxiv} 
\end{figure}

\begin{figure}[htb]
\centering
\includegraphics[width=1 \textwidth]{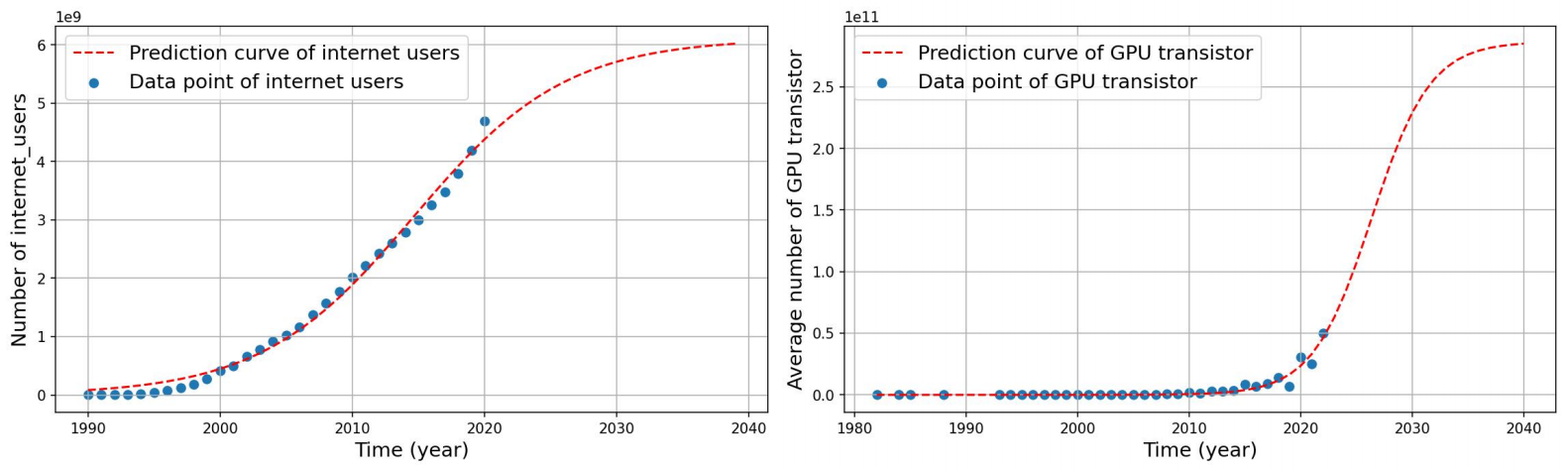}
\caption{The data and prediction results on the annual cumulative number of internet users and GPU transistors.}
\label{fig:data_gpu} 
\end{figure}

\subsection{Cross-Validation Experiments (RQ4)}
In order to verify our conclusions about the development of AI technology obtained from AI Historical Statistics dataset, we involve a dataset from another source for cross-validation. In this case, we employ AI Arxiv Papers dataset, which collects the annual cumulative number of papers on different topics related to AI technology on the arxiv website from 2008 to 2023. From the results in Table~\ref{tab:variance}, we obtain the $2\sigma$ time range of total AI development in the third wave is [2009, 2040]. Since the development of AI technology after 2008 is mainly in the third wave, we can use the ordinary logistic growth process on the AI Arxiv Papers dataset. 
The values of the optimal parameters and time ranges are shown in Table~\ref{tab:performance}, and the visualization of the fitting curves are shown in Figure~\ref{fig:arxiv}. From the fitting curves in Figure~\ref{fig:arxiv}, the logistic growth process can accurately characterize the development trend of the third AI wave. From the values of parameters in Table~\ref{tab:performance}, we can find that the midpoint of total development is around 2024, and it will fall into decline around 2035-2040, which is almost consistent with the results we predicted on AI Historical Statistics dataset. In addition, these experimental results can also provide an approximate prediction for different research fields related to AI technology.

In addition, we also analyze the relationship between the current AI wave and other limiting factors from the perspectives of data volume and computational facilities. As shown in Figure~\ref{fig:data_gpu}, we have perfectly fitted the annual growth trends of internet users and the average number of transistors in newly developed GPUs using a logistic curve. The prediction curves indicate that the growth of both internet users and GPU transistor counts will approach saturation after 2035, highlighting that the bottlenecks for data volume and computational resources needed for training large deep learning models will become increasingly apparent. Therefore, the development of next-generation AI technologies will require reducing dependency on data and further enhancing semiconductor design technologies.

\begin{table}[h]
  \caption{The optimal parameters and time range variables for each subject on the AI Arxiv Paper dataset.}
\centering
  \label{tab:performance}
  \scalebox{1.0}{
  \begin{tabular}{c|ccc|ccc}
	\toprule
			  & \multicolumn{3}{c}{Parameters} & \multicolumn{3}{c}{Time Range Variables}\tabularnewline
			  Subjects & $a$  & $w$ & $m$ & $\sigma$ & T(1$\sigma$) & T(2$\sigma$)\tabularnewline
		\midrule
    	Total & 388932.57  & 3.09 & 2024.88 & 5.60 & 2019-2030 & 2014-2036 \tabularnewline
		CV &82496.35	&2.29	&2022.03	&4.15	&2018-2026	&2014-2030 \tabularnewline
		CL &34420.36	&2.13	&2022.21	&3.87	&2018-2026	&2014-2030 \tabularnewline
		DA &9434.26	&3.08	&2018.92	&5.59	&2013-2025	&2008-2030\tabularnewline
		HCI &21523.80	&3.07	&2026.96	&5.57	&2021-2033	&2016-2038\tabularnewline
		SE &20282.21	&3.76	&2024.83	&6.83	&2018-2032	&2011-2038 \tabularnewline
		RO &33002.74	&2.51	&2025.08	&4.56	&2021-2030	&2016-2034 \tabularnewline
		CS &19133.25	&3.44	&2022.69	&6.24	&2016-2029	&2010-2035 \tabularnewline
		ML &54107.54	&2.14	&2025.32	&3.88	&2021-2029	&2018-2033 \tabularnewline
		\bottomrule
  \end{tabular}}
  \label{tab:performance}
\end{table}

\section{Discussion}
Different from some previous hypotheses positing exponential growth~\cite{kurzweil2005singularity}, we demonstrate that AI technology development conforms to a multi-logistic growth process through comprehensive experiments. Our research emphasizes the importance of grounding such discussions in the objective laws governing technological progress. we summarize the following two key insights as follows:

\textbf{The impact of the current AI wave on society still needs to be calmly evaluated.}
Although current AI technologies are undeniably reshaping various societal facets, a prudent assessment is crucial before speculative discussions about technological singularities can be substantiated. Despite the transformative effects of current AI advancements, particularly exemplified by LLMs, there is no conclusive evidence suggesting these technologies have initiated a new wave of AI development. Rather, LLMs appear as natural extensions within the continuum of deep learning evolution, leveraging advancements in deep network architectures and computing hardware. 
It is conceivable that LLMs and their applications may become as popular as smartphones in the future, but without theoretical breakthroughs that go beyond the current deep learning paradigm, they are still far from causing the kind of social changes described by ‘technological singularity’.

\textbf{The advancement of AI technology requires further breakthroughs in fundamental theories.}
The future trajectory of AI technology may be characterized by 'short-term pessimism but long-term optimism'. In the short term, concerns arise from potential data exhaustion in deep learning training\cite{villalobos2022will}, along with hallucination and security problems inherent in LLMs~\cite{hadi2023survey}. These challenges underscore the inefficiencies of the current deep learning paradigm, which struggles to align seamlessly with human societal needs.
In the long term, we expect new technological breakthroughs to emerge before the third AI wave is about expires, just like the previous two AI waves. Achieving such breakthroughs, however, necessitates fundamental theoretical advancements and may require comprehensive research in complexity theory and cognitive science within AI. By drawing inspiration from human brain function and elucidating the emergence mechanisms of LLMs, we can potentially address these limitations at the source.

\section{Conclusion and Future Direction}
In this paper, we effectively model the history of AI technology development through multi-logistic growth process and forecast its future trends. In the experiments, we not only demonstrate the rationality and superiority of our proposed model through comparative experiments, but also comprehensively analyze the important parameters and derivative characteristics of the fitted model. Based on the experimental results, we can characterize the current and predict future trends of AI technology: the third wave of AI develops most rapidly around 2024, but this wave seems likely to fade away around 2035-2040 if there is no further breakthrough in the underlying theories.
Therefore, the technological singularity may come in the distant future, but it will not come soon in the foreseeable future. Despite this, the impact of the current AI wave on the progress of human society cannot be ignored. At least AI can become a universal tool that brings convenience to humans, just like smartphones.

In the future, we will continue to monitor developments in AI technology and improved our model parameters annually based on new data. This iterative approach will ensure that our predictions remain accurate and relevant in the rapidly evolving field of AI research. In addition, we will also delve deeper into the development mechanism of AI technology from more essential perspectives, such as from the perspective of the development of computer semiconductor technology, database technology, etc.





\bibliographystyle{elsarticle-num-names}
\bibliography{sample.bib}

\begin{thebibliography}{46}
\expandafter\ifx\csname natexlab\endcsname\relax\def\natexlab#1{#1}\fi
\providecommand{\url}[1]{\texttt{#1}}
\providecommand{\href}[2]{#2}
\providecommand{\path}[1]{#1}
\providecommand{\DOIprefix}{doi:}
\providecommand{\ArXivprefix}{arXiv:}
\providecommand{\URLprefix}{URL: }
\providecommand{\Pubmedprefix}{pmid:}
\providecommand{\doi}[1]{\href{http://dx.doi.org/#1}{\path{#1}}}
\providecommand{\Pubmed}[1]{\href{pmid:#1}{\path{#1}}}
\providecommand{\bibinfo}[2]{#2}
\ifx\xfnm\relax \def\xfnm[#1]{\unskip,\space#1}\fi
\bibitem[{Fast and Horvitz(2017)}]{fast2017long}
\bibinfo{author}{E.~Fast}, \bibinfo{author}{E.~Horvitz},
\newblock \bibinfo{title}{Long-term trends in the public perception of artificial intelligence},
\newblock in: \bibinfo{booktitle}{Proceedings of the AAAI conference on artificial intelligence}, volume~\bibinfo{volume}{31}, \bibinfo{year}{2017}.
\bibitem[{Moor(2006)}]{moor2006dartmouth}
\bibinfo{author}{J.~Moor},
\newblock \bibinfo{title}{The dartmouth college artificial intelligence conference: The next fifty years},
\newblock \bibinfo{journal}{Ai Magazine} \bibinfo{volume}{27} (\bibinfo{year}{2006}) \bibinfo{pages}{87--87}.
\bibitem[{Wang et~al.(2021)Wang, Liu, Liu, and Tao}]{wang2021artificial}
\bibinfo{author}{L.~Wang}, \bibinfo{author}{Z.~Liu}, \bibinfo{author}{A.~Liu}, \bibinfo{author}{F.~Tao},
\newblock \bibinfo{title}{Artificial intelligence in product lifecycle management},
\newblock \bibinfo{journal}{The International Journal of Advanced Manufacturing Technology} \bibinfo{volume}{114} (\bibinfo{year}{2021}) \bibinfo{pages}{771--796}.
\bibitem[{Chang et~al.(2024)Chang, Wang, Wang, Wu, Yang, Zhu, Chen, Yi, Wang, Wang et~al.}]{chang2024survey}
\bibinfo{author}{Y.~Chang}, \bibinfo{author}{X.~Wang}, \bibinfo{author}{J.~Wang}, \bibinfo{author}{Y.~Wu}, \bibinfo{author}{L.~Yang}, \bibinfo{author}{K.~Zhu}, \bibinfo{author}{H.~Chen}, \bibinfo{author}{X.~Yi}, \bibinfo{author}{C.~Wang}, \bibinfo{author}{Y.~Wang}, et~al.,
\newblock \bibinfo{title}{A survey on evaluation of large language models},
\newblock \bibinfo{journal}{ACM Transactions on Intelligent Systems and Technology} \bibinfo{volume}{15} (\bibinfo{year}{2024}) \bibinfo{pages}{1--45}.
\bibitem[{Liu et~al.(2024)Liu, Zhang, Li, Yan, Gao, Chen, Yuan, Huang, Sun, Gao et~al.}]{liu2024sora}
\bibinfo{author}{Y.~Liu}, \bibinfo{author}{K.~Zhang}, \bibinfo{author}{Y.~Li}, \bibinfo{author}{Z.~Yan}, \bibinfo{author}{C.~Gao}, \bibinfo{author}{R.~Chen}, \bibinfo{author}{Z.~Yuan}, \bibinfo{author}{Y.~Huang}, \bibinfo{author}{H.~Sun}, \bibinfo{author}{J.~Gao}, et~al.,
\newblock \bibinfo{title}{Sora: A review on background, technology, limitations, and opportunities of large vision models},
\newblock \bibinfo{journal}{arXiv preprint arXiv:2402.17177}  (\bibinfo{year}{2024}).
\bibitem[{Liu et~al.(2023)Liu, Zheng, Du, Ding, Qian, Yang, and Tang}]{liu2023gpt}
\bibinfo{author}{X.~Liu}, \bibinfo{author}{Y.~Zheng}, \bibinfo{author}{Z.~Du}, \bibinfo{author}{M.~Ding}, \bibinfo{author}{Y.~Qian}, \bibinfo{author}{Z.~Yang}, \bibinfo{author}{J.~Tang},
\newblock \bibinfo{title}{Gpt understands, too},
\newblock \bibinfo{journal}{AI Open}  (\bibinfo{year}{2023}).
\bibitem[{Abd-Alrazaq et~al.(2023)Abd-Alrazaq, AlSaad, Alhuwail, Ahmed, Healy, Latifi, Aziz, Damseh, Alrazak, Sheikh et~al.}]{abd2023large}
\bibinfo{author}{A.~Abd-Alrazaq}, \bibinfo{author}{R.~AlSaad}, \bibinfo{author}{D.~Alhuwail}, \bibinfo{author}{A.~Ahmed}, \bibinfo{author}{P.~M. Healy}, \bibinfo{author}{S.~Latifi}, \bibinfo{author}{S.~Aziz}, \bibinfo{author}{R.~Damseh}, \bibinfo{author}{S.~A. Alrazak}, \bibinfo{author}{J.~Sheikh}, et~al.,
\newblock \bibinfo{title}{Large language models in medical education: opportunities, challenges, and future directions},
\newblock \bibinfo{journal}{JMIR Medical Education} \bibinfo{volume}{9} (\bibinfo{year}{2023}) \bibinfo{pages}{e48291}.
\bibitem[{Biever(2023)}]{biever2023chatgpt}
\bibinfo{author}{C.~Biever},
\newblock \bibinfo{title}{Chatgpt broke the turing test-the race is on for new ways to assess ai},
\newblock \bibinfo{journal}{Nature} \bibinfo{volume}{619} (\bibinfo{year}{2023}) \bibinfo{pages}{686--689}.
\bibitem[{Kurzweil(2005)}]{kurzweil2005singularity}
\bibinfo{author}{R.~Kurzweil},
\newblock \bibinfo{title}{The singularity is near},
\newblock in: \bibinfo{booktitle}{Ethics and emerging technologies}, \bibinfo{publisher}{Springer}, \bibinfo{year}{2005}, pp. \bibinfo{pages}{393--406}.
\bibitem[{Vaswani et~al.(2017)Vaswani, Shazeer, Parmar, Uszkoreit, Jones, Gomez, Kaiser, and Polosukhin}]{vaswani2017attention}
\bibinfo{author}{A.~Vaswani}, \bibinfo{author}{N.~Shazeer}, \bibinfo{author}{N.~Parmar}, \bibinfo{author}{J.~Uszkoreit}, \bibinfo{author}{L.~Jones}, \bibinfo{author}{A.~N. Gomez}, \bibinfo{author}{{\L}.~Kaiser}, \bibinfo{author}{I.~Polosukhin},
\newblock \bibinfo{title}{Attention is all you need},
\newblock \bibinfo{journal}{Advances in neural information processing systems} \bibinfo{volume}{30} (\bibinfo{year}{2017}).
\bibitem[{Kenton and Toutanova(2019)}]{kenton2019bert}
\bibinfo{author}{J.~D. M.-W.~C. Kenton}, \bibinfo{author}{L.~K. Toutanova},
\newblock \bibinfo{title}{Bert: Pre-training of deep bidirectional transformers for language understanding},
\newblock in: \bibinfo{booktitle}{Proceedings of NAACL-HLT}, \bibinfo{year}{2019}, pp. \bibinfo{pages}{4171--4186}.
\bibitem[{Hadi et~al.(2023)Hadi, Qureshi, Shah, Irfan, Zafar, Shaikh, Akhtar, Wu, Mirjalili et~al.}]{hadi2023survey}
\bibinfo{author}{M.~U. Hadi}, \bibinfo{author}{R.~Qureshi}, \bibinfo{author}{A.~Shah}, \bibinfo{author}{M.~Irfan}, \bibinfo{author}{A.~Zafar}, \bibinfo{author}{M.~B. Shaikh}, \bibinfo{author}{N.~Akhtar}, \bibinfo{author}{J.~Wu}, \bibinfo{author}{S.~Mirjalili}, et~al.,
\newblock \bibinfo{title}{A survey on large language models: Applications, challenges, limitations, and practical usage},
\newblock \bibinfo{journal}{Authorea Preprints}  (\bibinfo{year}{2023}).
\bibitem[{Jin et~al.(2023{\natexlab{a}})Jin, Liang, Fang, Shao, Huang, Zhang, and Zheng}]{jin2023spatio}
\bibinfo{author}{G.~Jin}, \bibinfo{author}{Y.~Liang}, \bibinfo{author}{Y.~Fang}, \bibinfo{author}{Z.~Shao}, \bibinfo{author}{J.~Huang}, \bibinfo{author}{J.~Zhang}, \bibinfo{author}{Y.~Zheng},
\newblock \bibinfo{title}{Spatio-temporal graph neural networks for predictive learning in urban computing: A survey},
\newblock \bibinfo{journal}{IEEE Transactions on Knowledge and Data Engineering}  (\bibinfo{year}{2023}{\natexlab{a}}).
\bibitem[{Jin et~al.(2023{\natexlab{b}})Jin, Liu, Li, and Huang}]{jin2023con}
\bibinfo{author}{G.~Jin}, \bibinfo{author}{L.~Liu}, \bibinfo{author}{F.~Li}, \bibinfo{author}{J.~Huang},
\newblock \bibinfo{title}{Spatio-temporal graph neural point process for traffic congestion event prediction},
\newblock in: \bibinfo{booktitle}{Proceedings of the AAAI Conference on Artificial Intelligence}, volume~\bibinfo{volume}{37}, \bibinfo{year}{2023}{\natexlab{b}}, pp. \bibinfo{pages}{14268--14276}.
\bibitem[{Good(1966)}]{good1966speculations}
\bibinfo{author}{I.~J. Good},
\newblock \bibinfo{title}{Speculations concerning the first ultraintelligent machine},
\newblock in: \bibinfo{booktitle}{Advances in computers}, volume~\bibinfo{volume}{6}, \bibinfo{publisher}{Elsevier}, \bibinfo{year}{1966}, pp. \bibinfo{pages}{31--88}.
\bibitem[{Vinge(1993)}]{vinge1993coming}
\bibinfo{author}{V.~Vinge},
\newblock \bibinfo{title}{The coming technological singularity: How to survive in the post-human era},
\newblock \bibinfo{journal}{Science fiction criticism: An anthology of essential writings} \bibinfo{volume}{81} (\bibinfo{year}{1993}) \bibinfo{pages}{352--363}.
\bibitem[{Yudkowsky(1996)}]{yudkowsky1996staring}
\bibinfo{author}{E.~Yudkowsky},
\newblock \bibinfo{title}{Staring into the singularity}  (\bibinfo{year}{1996}).
\bibitem[{Grace et~al.(2018)Grace, Salvatier, Dafoe, Zhang, and Evans}]{grace2018will}
\bibinfo{author}{K.~Grace}, \bibinfo{author}{J.~Salvatier}, \bibinfo{author}{A.~Dafoe}, \bibinfo{author}{B.~Zhang}, \bibinfo{author}{O.~Evans},
\newblock \bibinfo{title}{When will ai exceed human performance? evidence from ai experts},
\newblock \bibinfo{journal}{Journal of Artificial Intelligence Research} \bibinfo{volume}{62} (\bibinfo{year}{2018}) \bibinfo{pages}{729--754}.
\bibitem[{Harris et~al.(2018)Harris, Devkota, Khanna, Eranki, and Landis}]{harris2018logistic}
\bibinfo{author}{T.~M. Harris}, \bibinfo{author}{J.~P. Devkota}, \bibinfo{author}{V.~Khanna}, \bibinfo{author}{P.~L. Eranki}, \bibinfo{author}{A.~E. Landis},
\newblock \bibinfo{title}{Logistic growth curve modeling of us energy production and consumption},
\newblock \bibinfo{journal}{Renewable and Sustainable Energy Reviews} \bibinfo{volume}{96} (\bibinfo{year}{2018}) \bibinfo{pages}{46--57}.
\bibitem[{Burg and Ausubel(2021)}]{burg2021moore}
\bibinfo{author}{D.~Burg}, \bibinfo{author}{J.~H. Ausubel},
\newblock \bibinfo{title}{Moore’s law revisited through intel chip density},
\newblock \bibinfo{journal}{PloS one} \bibinfo{volume}{16} (\bibinfo{year}{2021}) \bibinfo{pages}{e0256245}.
\bibitem[{Shao et~al.(2022)Shao, Zhao, Yuan, Ding, and Wang}]{shao2022tracing}
\bibinfo{author}{Z.~Shao}, \bibinfo{author}{R.~Zhao}, \bibinfo{author}{S.~Yuan}, \bibinfo{author}{M.~Ding}, \bibinfo{author}{Y.~Wang},
\newblock \bibinfo{title}{Tracing the evolution of ai in the past decade and forecasting the emerging trends},
\newblock \bibinfo{journal}{Expert Systems with Applications} \bibinfo{volume}{209} (\bibinfo{year}{2022}) \bibinfo{pages}{118221}.
\bibitem[{Dwivedi et~al.(2023)Dwivedi, Sharma, Rana, Giannakis, Goel, and Dutot}]{dwivedi2023evolution}
\bibinfo{author}{Y.~K. Dwivedi}, \bibinfo{author}{A.~Sharma}, \bibinfo{author}{N.~P. Rana}, \bibinfo{author}{M.~Giannakis}, \bibinfo{author}{P.~Goel}, \bibinfo{author}{V.~Dutot},
\newblock \bibinfo{title}{Evolution of artificial intelligence research in technological forecasting and social change: Research topics, trends, and future directions},
\newblock \bibinfo{journal}{Technological Forecasting and Social Change} \bibinfo{volume}{192} (\bibinfo{year}{2023}) \bibinfo{pages}{122579}.
\bibitem[{Baum et~al.(2011)Baum, Goertzel, and Goertzel}]{baum2011long}
\bibinfo{author}{S.~D. Baum}, \bibinfo{author}{B.~Goertzel}, \bibinfo{author}{T.~G. Goertzel},
\newblock \bibinfo{title}{How long until human-level ai? results from an expert assessment},
\newblock \bibinfo{journal}{Technological Forecasting and Social Change} \bibinfo{volume}{78} (\bibinfo{year}{2011}) \bibinfo{pages}{185--195}.
\bibitem[{Gruetzemacher et~al.(2021)Gruetzemacher, Dorner, Bernaola-Alvarez, Giattino, and Manheim}]{gruetzemacher2021forecasting}
\bibinfo{author}{R.~Gruetzemacher}, \bibinfo{author}{F.~E. Dorner}, \bibinfo{author}{N.~Bernaola-Alvarez}, \bibinfo{author}{C.~Giattino}, \bibinfo{author}{D.~Manheim},
\newblock \bibinfo{title}{Forecasting ai progress: A research agenda},
\newblock \bibinfo{journal}{Technological Forecasting and Social Change} \bibinfo{volume}{170} (\bibinfo{year}{2021}) \bibinfo{pages}{120909}.
\bibitem[{Halal(2013)}]{halal2013forecasting}
\bibinfo{author}{W.~E. Halal},
\newblock \bibinfo{title}{Forecasting the technology revolution: results and learnings from the techcast project},
\newblock \bibinfo{journal}{Technological Forecasting and Social Change} \bibinfo{volume}{80} (\bibinfo{year}{2013}) \bibinfo{pages}{1635--1643}.
\bibitem[{Villalobos et~al.(2022)Villalobos, Sevilla, Heim, Besiroglu, Hobbhahn, and Ho}]{villalobos2022will}
\bibinfo{author}{P.~Villalobos}, \bibinfo{author}{J.~Sevilla}, \bibinfo{author}{L.~Heim}, \bibinfo{author}{T.~Besiroglu}, \bibinfo{author}{M.~Hobbhahn}, \bibinfo{author}{A.~Ho},
\newblock \bibinfo{title}{Will we run out of data? an analysis of the limits of scaling datasets in machine learning},
\newblock \bibinfo{journal}{arXiv preprint arXiv:2211.04325}  (\bibinfo{year}{2022}).
\bibitem[{Modis(2022)}]{modis2022links}
\bibinfo{author}{T.~Modis},
\newblock \bibinfo{title}{Links between entropy, complexity, and the technological singularity},
\newblock \bibinfo{journal}{Technological Forecasting and Social Change} \bibinfo{volume}{176} (\bibinfo{year}{2022}) \bibinfo{pages}{121457}.
\bibitem[{Verhulst(1845)}]{verhulst1845recherches}
\bibinfo{author}{P.-F. Verhulst},
\newblock \bibinfo{title}{Recherches math{\'e}matiques sur la loi d’accroissement de la population},
\newblock \bibinfo{journal}{M{\'e}moires de l'Acad{\'e}mie royale de Belgique} \bibinfo{volume}{18} (\bibinfo{year}{1845}) \bibinfo{pages}{1--40}.
\bibitem[{McKendrick and Pai(1912)}]{mckendrick1912xlv}
\bibinfo{author}{A.~McKendrick}, \bibinfo{author}{M.~K. Pai},
\newblock \bibinfo{title}{Xlv.—the rate of multiplication of micro-organisms: a mathematical study},
\newblock \bibinfo{journal}{Proceedings of the Royal Society of Edinburgh} \bibinfo{volume}{31} (\bibinfo{year}{1912}) \bibinfo{pages}{649--655}.
\bibitem[{Sepaskhah et~al.(2011)Sepaskhah, Fahandezh-Saadi, and Zand-Parsa}]{sepaskhah2011logistic}
\bibinfo{author}{A.~R. Sepaskhah}, \bibinfo{author}{S.~Fahandezh-Saadi}, \bibinfo{author}{S.~Zand-Parsa},
\newblock \bibinfo{title}{Logistic model application for prediction of maize yield under water and nitrogen management},
\newblock \bibinfo{journal}{Agricultural Water Management} \bibinfo{volume}{99} (\bibinfo{year}{2011}) \bibinfo{pages}{51--57}.
\bibitem[{Laird(1964)}]{laird1964dynamics}
\bibinfo{author}{A.~K. Laird},
\newblock \bibinfo{title}{Dynamics of tumour growth},
\newblock \bibinfo{journal}{British journal of cancer} \bibinfo{volume}{18} (\bibinfo{year}{1964}) \bibinfo{pages}{490}.
\bibitem[{Kirkwood(2015)}]{kirkwood2015deciphering}
\bibinfo{author}{T.~B. Kirkwood},
\newblock \bibinfo{title}{Deciphering death: a commentary on gompertz (1825)‘on the nature of the function expressive of the law of human mortality, and on a new mode of determining the value of life contingencies’},
\newblock \bibinfo{journal}{Philosophical Transactions of the Royal Society B: Biological Sciences} \bibinfo{volume}{370} (\bibinfo{year}{2015}) \bibinfo{pages}{20140379}.
\bibitem[{Richards(1959)}]{richards1959flexible}
\bibinfo{author}{F.~J. Richards},
\newblock \bibinfo{title}{A flexible growth function for empirical use},
\newblock \bibinfo{journal}{Journal of experimental Botany} \bibinfo{volume}{10} (\bibinfo{year}{1959}) \bibinfo{pages}{290--301}.
\bibitem[{Lee et~al.(2020)Lee, Lei, and Mallick}]{lee2020estimation}
\bibinfo{author}{S.~Y. Lee}, \bibinfo{author}{B.~Lei}, \bibinfo{author}{B.~Mallick},
\newblock \bibinfo{title}{Estimation of covid-19 spread curves integrating global data and borrowing information},
\newblock \bibinfo{journal}{PloS one} \bibinfo{volume}{15} (\bibinfo{year}{2020}) \bibinfo{pages}{e0236860}.
\bibitem[{Levene(2022)}]{levene2022skew}
\bibinfo{author}{M.~Levene},
\newblock \bibinfo{title}{A skew logistic distribution for modelling covid-19 waves and its evaluation using the empirical survival jensen--shannon divergence},
\newblock \bibinfo{journal}{Entropy} \bibinfo{volume}{24} (\bibinfo{year}{2022}) \bibinfo{pages}{600}.
\bibitem[{Perez(2003)}]{perez2003technological}
\bibinfo{author}{C.~Perez}, \bibinfo{title}{Technological revolutions and financial capital}, \bibinfo{publisher}{Edward Elgar Publishing}, \bibinfo{year}{2003}.
\bibitem[{Grubler(1990)}]{grubler1990rise}
\bibinfo{author}{A.~Grubler}, \bibinfo{title}{The rise and fall of infrastructures: dynamics of evolution and technological change in transport}, \bibinfo{publisher}{Physica-Verlag}, \bibinfo{year}{1990}.
\bibitem[{Kalem et~al.(2021)Kalem, Vayvay, Sennaroglu, and Tozan}]{kalem2021technology}
\bibinfo{author}{G.~Kalem}, \bibinfo{author}{O.~Vayvay}, \bibinfo{author}{B.~Sennaroglu}, \bibinfo{author}{H.~Tozan},
\newblock \bibinfo{title}{Technology forecasting in the mobile telecommunication industry: A case study towards the 5g era},
\newblock \bibinfo{journal}{Engineering Management Journal} \bibinfo{volume}{33} (\bibinfo{year}{2021}) \bibinfo{pages}{15--29}.
\bibitem[{Kucharavy and De~Guio(2011)}]{kucharavy2011logistic}
\bibinfo{author}{D.~Kucharavy}, \bibinfo{author}{R.~De~Guio},
\newblock \bibinfo{title}{Logistic substitution model and technological forecasting},
\newblock \bibinfo{journal}{Procedia Engineering} \bibinfo{volume}{9} (\bibinfo{year}{2011}) \bibinfo{pages}{402--416}.
\bibitem[{Fletcher(1971)}]{fletcher1971modified}
\bibinfo{author}{R.~Fletcher},
\newblock \bibinfo{title}{A modified marquardt subroutine for non-linear least squares}  (\bibinfo{year}{1971}).
\bibitem[{Dro{\.z}d{\.z} et~al.(2003)Dro{\.z}d{\.z}, Gr{\"u}mmer, Ruf, and Speth}]{drozdz2003log}
\bibinfo{author}{S.~Dro{\.z}d{\.z}}, \bibinfo{author}{F.~Gr{\"u}mmer}, \bibinfo{author}{F.~Ruf}, \bibinfo{author}{J.~Speth},
\newblock \bibinfo{title}{Log-periodic self-similarity: an emerging financial law?},
\newblock \bibinfo{journal}{Physica A: Statistical Mechanics and its Applications} \bibinfo{volume}{324} (\bibinfo{year}{2003}) \bibinfo{pages}{174--182}.
\bibitem[{Kurzweil(2001)}]{kurzweil2001law}
\bibinfo{author}{R.~Kurzweil},
\newblock \bibinfo{title}{The law of accelerating returns},
\newblock in: \bibinfo{booktitle}{Alan Turing: Life and legacy of a great thinker}, \bibinfo{publisher}{Springer}, \bibinfo{year}{2001}, pp. \bibinfo{pages}{381--416}.
\bibitem[{Rumelhart et~al.(1986)Rumelhart, Hinton, and Williams}]{rumelhart1986learning}
\bibinfo{author}{D.~E. Rumelhart}, \bibinfo{author}{G.~E. Hinton}, \bibinfo{author}{R.~J. Williams},
\newblock \bibinfo{title}{Learning representations by back-propagating errors},
\newblock \bibinfo{journal}{nature} \bibinfo{volume}{323} (\bibinfo{year}{1986}) \bibinfo{pages}{533--536}.
\bibitem[{Hochreiter and Schmidhuber(1997)}]{hochreiter1997long}
\bibinfo{author}{S.~Hochreiter}, \bibinfo{author}{J.~Schmidhuber},
\newblock \bibinfo{title}{Long short-term memory},
\newblock \bibinfo{journal}{Neural computation} \bibinfo{volume}{9} (\bibinfo{year}{1997}) \bibinfo{pages}{1735--1780}.
\bibitem[{LeCun et~al.(1995)LeCun, Bengio et~al.}]{lecun1995convolutional}
\bibinfo{author}{Y.~LeCun}, \bibinfo{author}{Y.~Bengio}, et~al.,
\newblock \bibinfo{title}{Convolutional networks for images, speech, and time series},
\newblock \bibinfo{journal}{The handbook of brain theory and neural networks} \bibinfo{volume}{3361} (\bibinfo{year}{1995}) \bibinfo{pages}{1995}.
\bibitem[{Hsu(1999)}]{hsu1999ibm}
\bibinfo{author}{F.-h. Hsu},
\newblock \bibinfo{title}{Ibm's deep blue chess grandmaster chips},
\newblock \bibinfo{journal}{IEEE micro} \bibinfo{volume}{19} (\bibinfo{year}{1999}) \bibinfo{pages}{70--81}.

\end{thebibliography}







\end{document}